\documentclass[preprint,12pt]{elsarticle}

\usepackage{amssymb}
\usepackage{amsmath}
\usepackage{multirow}

\journal{Journal}

\begin{document}

\begin{frontmatter}



\title{Extended defects-enhanced oxygen diffusion in ThO$_2$}
\date{}
\author[inst1]{Miaomiao Jin\corref{cor1}}
\ead{mmjin@psu.edu}
\cortext[cor1]{Corresponding author}
\affiliation[inst1]{Department of Nuclear Engineering, The Pennsylvania State University, University Park, 16802 PA, USA}
\author[inst1]{Jilang Miao}
\author[inst1]{Beihan Chen}
\author[inst2]{Marat Khafizov}%
\affiliation[inst2]{Department of Mechanical and Aerospace Engineering, The Ohio State University, 201 W 19th Ave, Columbus, OH 43210, USA}
\author[inst3]{Yongfeng Zhang}%
\affiliation[inst3]{Department of Engineering Physics, University of Wisconsin-Madison, 1500 Engineering Dr, Madison, WI 53706, USA} 
\author[inst4]{David H. Hurley}%
\affiliation[inst4]{Idaho National Laboratory, 2525 Fremont Ave, Idaho Falls, ID 83402, USA}%

\begin{abstract}
Oxygen self-diffusion is key to understanding stoichiometry and defect structures in oxide nuclear fuels. Experimentally, low activation-barrier oxygen migration was found in ThO$_2$, a candidate nuclear fuel, possibly due to short-circuit diffusion mechanisms. Here, we perform extensive molecular dynamics simulations to show that various types of extended defects can enhance oxygen self-diffusion with a much-reduced activation barrier in ThO$_2$. In this work, we consider extended defects including 1D (dislocation), 2D (grain boundary), and 3D (void) defects. Due to the distinct characteristics of each type of extended defect, the modulation of oxygen diffusion varies. These results provide a quantitative description of oxygen transport, which is significantly enhanced within a close distance (nanometer scale) from the extended defects. Among all these defects, grain boundary, particularly the $\Sigma 3$ twin boundary with a low formation energy, exhibits the strongest effect on increasing oxygen transport.  
\end{abstract}


\end{frontmatter}


\section{Introduction}
\label{sec:sample1}
Oxygen transport in oxide nuclear fuels is of significance to understanding the local stoichiometry, phase stability, and oxidation behavior. Due to the high oxygen diffusivity compared to the cations, the ionic conductivity is dominated by oxygen atoms. Under radiation environments, a variety of defects are formed and evolving in nuclear fuels, which affects oxygen transport. There have been significant efforts in quantifying the mass transport properties of fuel cation, oxygen, and fission products in nuclear fuels, particularly in UO$_2$ (see \cite{murch1987oxygen,williams2015atomistic} and references within), given its practical importance in nuclear reactors. In comparison, ThO$_2$ as a candidate nuclear fuel has received much less attention on oxygen diffusion behavior, particularly how oxygen diffusion is affected by extended defects. The most recent computational results are based on the bulk crystal \cite{cooper2015modeling}, investigating how the pressure factors into the diffusion coefficients. Taking into account the effect of grain boundaries and dislocations, oxygen diffusion in pure UO$_2$ has been evaluated with molecular dynamics (MD) \cite{williams2015atomistic,murphy2014pipe,vincent2009self,arima2010molecular}, where stronger oxygen diffusion at these extended defects was identified than that in the bulk region. It is not clear whether the conclusions from UO$_2$ defective systems would be transferable to ThO$_2$.         
 
In ThO$_2$, oxygen self-diffusion coefficients in the bulk have been experimentally measured in previous studies, and a broad range of Arrhenius energies was reported. Edward et al. \cite{ando1981oxygen,colbourn1983calculated} proposed 2.85 eV with data from 800-1600 $\mathrm{^oC}$, based on the interpretation of intrinsic diffusion mechanisms (involving the formation of an oxygen Frenkel pair), while later experiments suggested 1.80-1.92 eV \cite{lee1973electrical,choudhury1974transition}. In more recent results by Ando et al. \cite{ando1981oxygen}, intrinsic oxygen diffusion with an activation energy of 2.16 eV was used to account for the high-temperature regime, and extrinsic diffusion, presumably due to dislocation-enhanced diffusion with activation energy 0.76 eV was used to explain the low-temperature regime. From the theoretical perspective, the self-diffusion activation energy needs to incorporate both oxygen Frenkel pair formation and migration. Later, Colbourn and Mackrodt \cite{colbourn1983calculated} reassessed the reported diffusion data based on electrical conductivity and argued that i) these values ($\sim$ 2 eV) are too small to account for the intrinsic diffusion of oxygen, as the formation energy of oxygen Frenkel is relatively high (5.89-9.8 eV from computational studies \cite{colbourn1983calculated,cooper2014many,yun2009defect}) and ii) the diffusing oxygen species is not determined. In the same work \cite{colbourn1983calculated}, it was suggested that the low Arrhenius energy by Ando et al. \cite{ando1981oxygen} involves the creation of oxygen interstitials ($\mathrm{O}_i^{2-}$) and holes in the valence band, while the value by Edward et al. \cite{ando1981oxygen,colbourn1983calculated} involves a different oxygen vacancy formation mechanism. These understandings of the bulk behavior could shed light on the oxygen transport near extended defects. 

Short-circuit diffusion via extended defects has long been recognized in crystalline systems. Due to the specific features in defects such as dislocations and interfaces, short-circuit diffusion mechanisms can arise, and contribute significantly to low-temperature mass transport. For example, in oxides such as urania \cite{murphy2014pipe}, magnesia \cite{zhang2010defects}, and alumina \cite{tang2003determination}, dislocations are known as fast diffusion paths due to the low coordination. Such pipe diffusion can contribute significantly to  the ionic conductivity at low temperatures. It should be noted that dislocation pipe-diffusion may not always occur, e.g., in doped CeO$_2$, dislocations were found to slow down oxygen diffusion due to charge effects upon elemental segregation \cite{sun2015edge}. With polycrystalline samples, at GB regions, due to differences in crystallographic orientations and atomic arrangements between adjacent grains, GBs can act as fast diffusion paths as reported in UO$_2$ \cite{williams2015atomistic,vincent2009self,arima2010molecular}, and the impact was shown to depend on the specific character of GB. Finally, the free surfaces, due to reduced bonding, would enable efficient atom jumps, and this applicability may also extend to curved surfaces such as voids and bubbles, which tend to form in oxide fuels during fabrication and normal operation stages \cite{whapham1966electron}.

In this work, these generally recognized fast diffusion pathways are systematically examined in ThO$_2$ to gain an understanding of their impact on oxygen diffusion under 3000 K, where the superionic transition is yet to occur \cite{ghosh2016computational}. Here, we quantify the impact of individual extended defects including 1D dislocation, 2D grain boundary, and 3D voids. The oxygen diffusion is found to be enhanced in all the considered cases, but the boosting factor differs due to the unique characteristics of each defect. As irradiated fuels are expected to have a high concentration of multi-dimensional defects, particularly in the high-burnup structure where grain refinement can occur, such quantification provides a basis for evaluating the effective oxygen diffusion in ThO$_2$.

\section{Methods}
Classical molecular dynamics (MD) approach as implemented in LAMMPS \cite{plimpton2007lammps} is utilized. Atom interactions are described by an embedded-atom potential developed by Cooper et al. \cite{cooper2014many} for ThO$_2$. In this potential, both Th (+2.2208) and O (-1.1104) atoms are modeled with fixed non-formal charges, and this potential can well reproduce thermophysical and defect properties \cite{cooper2014many,cooper2015modellingthermal,cooper2015thermophysical}. The simulation cells for each type of defect are adapted to the specific structure to be detailed below. In all structures constructed, the whole cell is charge neutral to avoid numerical problems from periodic boundary conditions. The range of temperature in simulations is from 2000 K to 3000 K where sufficient atomic displacements occur within the MD timescale to yield confident statistics. Here we limit the considered temperature range below 3000 K as it was found that nonlinearity in the oxygen diffusion coefficient ($D_{\mathrm{O}}$) appears in the ``superionic" regime at a temperature close to but below the melting point due to dynamic disorder in the oxygen sublattice. In ThO$_2$, MD simulations show that the superionic transition occurred at 3000 K \cite{ghosh2016computational}. After the structure is constructed, the corresponding cell is first fully relaxed at zero pressure at a specific temperature (NPT), and then atomic snapshots are taken from a canonical ensemble (NVT) run for 200-500 ps to compute the mean squared displacement (MSD). The time step is set to 1.0 fs. For the statistical quality of results, eight independent simulations are run for each type of structure at each temperature. Finally, the diffusion coefficients are computed based on Einstein diffusion analysis, i.e., 
\begin{equation}
\label{eq:D}
D_{\mathrm{O},\alpha} = \lim\limits_{t \to \infty}\frac{\langle(\mathbf{r}_{\mathrm{O},\alpha}(t)-\mathbf{r}_{\mathrm{O},\alpha}(0)\rangle}{2t}
\end{equation}
where $\alpha$ stands for the direction ($x$, $y$, and $z$), $\mathbf{r}_{\mathrm{O},\alpha}(t)$ and $\mathbf{r}_{\mathrm{O},\alpha}(0)$ are the oxygen positions at time $t$ and 0, respectively. Practically, linear regression between MSD and $t$ is performed to reduce uncertainty in $D_{\mathrm{O},\alpha}$. By distinguishing each direction, we can identify any low-dimensional oxygen diffusion pattern pertinent to the defect structure. To obtain spatial dependence of $D_{\mathrm{O},\alpha}$, the atoms are assigned to ``slabs"/``shells" at $t=0$, and Eq. \ref{eq:D} is applied to analyze each slab/shell of oxygen atoms. This method has also been used in previous studies of oxygen diffusion in UO$_2$ under the influence of dislocations and GBs \cite{williams2015atomistic,murphy2014pipe}.  

\noindent\textbf{Dislocation}
 
\begin{figure}[!ht]
	\centering
	\includegraphics[width=0.75\textwidth]{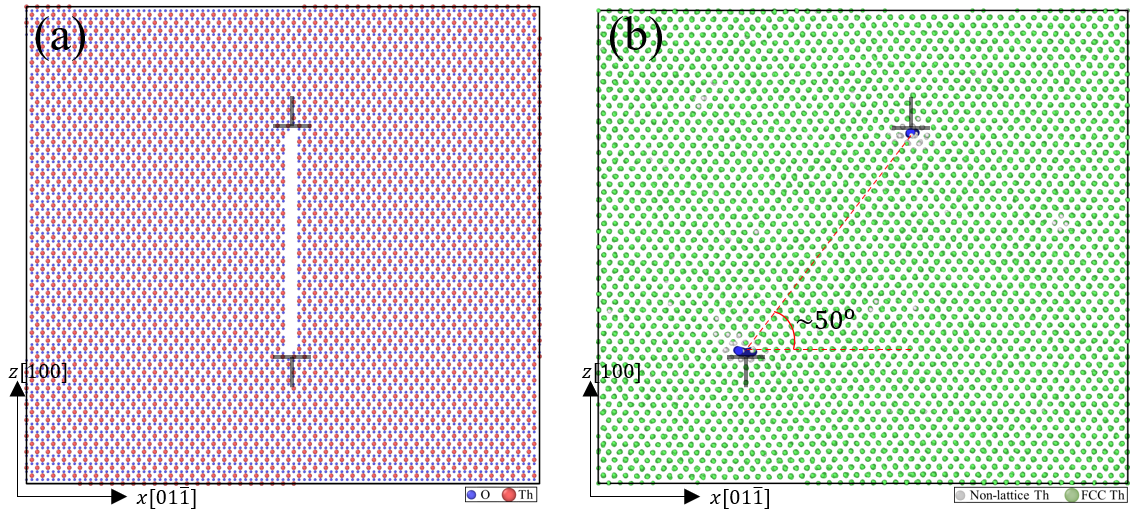}
	\caption{Dislocation construction by creating a slit in the simulation cell and subsequent high-temperature relaxation to obtain stable relative positions.}
	\label{fig:dislocation}
\end{figure}

Edge dislocation of Burgers vector $\mathbf{b}=1/2\langle110\rangle$ is considered, which is commonly seen in face-centered cubic (FCC) structures. To construct the dislocation, we first generate a perfect supercell containing 103,680 atoms ($\sim18.9\times4.7\times16.7$ nm$^3$) with the axes orientation as $x\,[01\bar{1}]$,  $y\,[011]$, and  $z\,[100]$ (Figure \ref{fig:dislocation}a). Then a slit is cut in the center along the $y$ -axis, and the ratio of removed O to Th atoms is 2 to ensure charge neutrality. To heal the structure, the system is relaxed at 2,500 K for 200 ps NPT and 1 ns NVT, leading to stable relative dislocation positions. Based on theoretical analysis, since the two dislocations' motion is limited to the slide plane, they should achieve equilibrium positions at 45$\mathrm{^o}$ \cite{hull2001introduction}. However, this degree is not strictly obeyed here due to limited cell size with periodic boundary conditions. In this case, we note that the high-temperature dynamics simulation leads to a relative position at $\sim 50 \mathrm{^o}$ (Figure \ref{fig:dislocation}b). This equilibration step is critical in the following MSD evaluation as moving dislocation poses a problem in quantifying the real MSD of atoms due to pipe diffusion, which has been noted in the previous study of UO$_2$ \cite{murphy2014pipe}. To obtain spatial dependence of $D_{\mathrm{O},\alpha}$, a cylindrical constant shell thickness of 3.0 {\AA} is used, with the origin at the dislocation core. 

\noindent\textbf{Grain boundary}
 
\begin{figure}[!ht]
	\centering
	\includegraphics[width=0.75\textwidth]{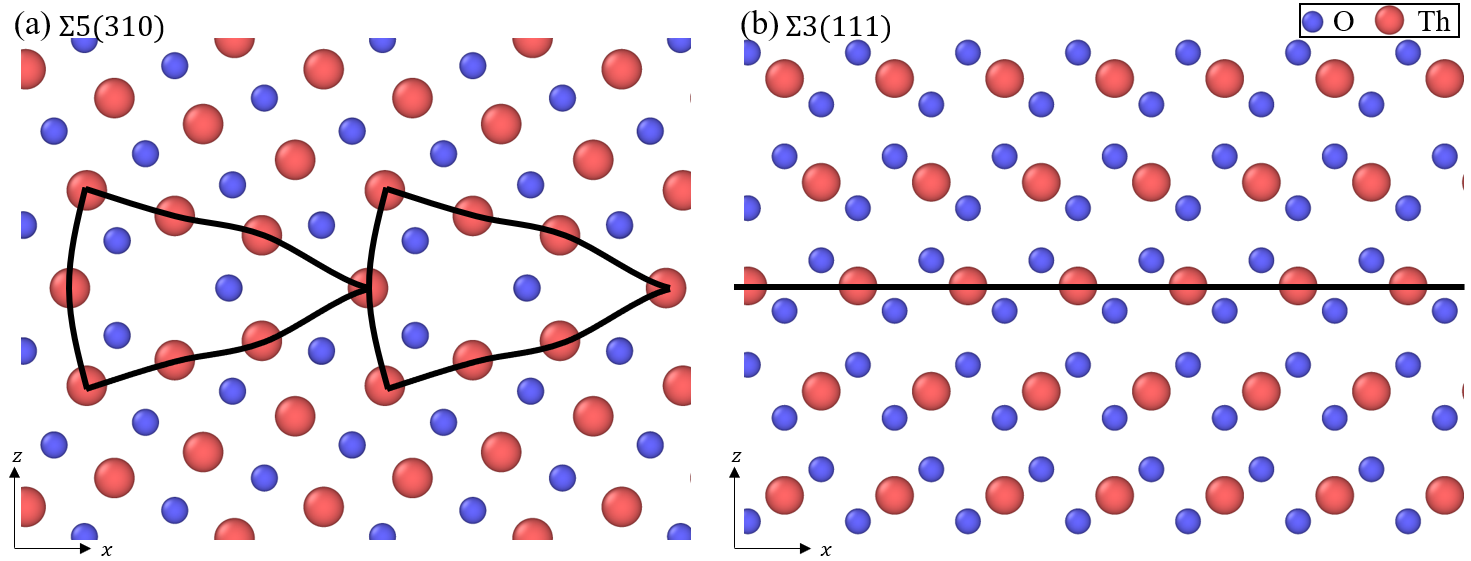}
	\caption{Energy-minimized GB structures: (a) $\Sigma 5(310)$ and (b) $\Sigma 3(111)$.}
	\label{fig:GB}
\end{figure}

In this study, we consider two representative types of symmetrically tilt grain boundaries, $\Sigma 5(310)/[001] $ and $\Sigma 3(111)/[\bar{1}10]$. The energy-minimized structures are shown in Figure \ref{fig:GB}, which are consistent with previous GB study in UO$_2$ \cite{williams2015atomistic}. The former has a large open space in GB, where strong pipe diffusion may occur, while the latter represents the most compact interface, which may differ from open-structured GBs in oxygen transport. The comparison between them can assist in interpreting the  diffusion behaviors for general types of GBs. Bi-crystal simulation cells are constructed. For $\Sigma 5(310)$, the simulation cell contains 57,600 atoms in $3.5\times3.3\times70.7$ nm$^3$, and for $\Sigma 3(111)$, it contains 92,160 atoms in $3.2\times5.5\times77.4$ nm$^3$ (the volumes change slightly at different temperatures). Note the distance between two GBs in the bi-crystal is intentionally made large to minimize GB-GB interaction. For the spatial dependence of $D_{\mathrm{O},\alpha}$, an increasing slab thickness from 3.0 {\AA} is used, with the origin at the GB.  

\noindent\textbf{Void}
 
Here, the curved surface is considered in the context of voids. The sources of curved surfaces in nuclear fuels could be from i) irradiation-induced voids, which are generally small in the nanometer range \cite{sabathier2014situ}, ii) the fabrication process, where $95\%$ porosity is commonly achieved with pore size on the $\mu m$-scale \cite{wangle2020morphology}, and iii) fission gas bubbles \cite{turnbull1971distribution}. In the current study, two void sizes (radius $r$ of 0.5 nm and 1.5 nm) are considered to reveal the difference in oxygen self-diffusion due to local curvature. A perfect system containing 96,000 atoms ($11\times11\times11$ nm$^3$) is first created; then for $r = 5$ \AA, a total of 36 atoms (12 Th and 24 O atoms) are removed, and for $r = 15$ \AA, 975 atoms (325 Th and 650 O atoms) are removed to form a void in the center. For spatial dependence of $D_{\mathrm{O},\alpha}$, a spherical constant shell thickness of 1.5 {\AA} is used, with the origin at the void center.

\section{Results and Discussion}
\begin{figure}[!ht]
	\centering
	\includegraphics[width=0.55\textwidth]{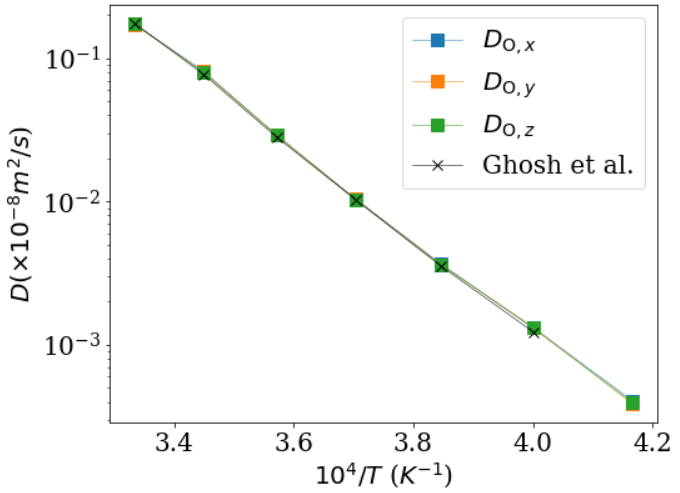}
	\caption{Oxygen diffusion coefficients as a function of the reciprocal of temperature. Data from Ghosh et al. \cite{ghosh2016computational} are plotted as a reference.}
	\label{fig:D_perfect}
\end{figure}

As a validation of the simulation settings, we first calculate the oxygen self-diffusion in a perfect lattice and compare it with literature values. Using a supercell containing 10$\times$10$\times$10 unit cells, the calculation is performed within the range of 2400 K to 3000 K with increments of 100 K. Then Eq. \ref{eq:D} is applied to extract $D_\mathrm{O}$. As shown in Figure \ref{fig:D_perfect}, the self-diffusion along each axis overlaps well with the previous bulk calculation using the same potential \cite{ghosh2016computational}. By fitting the data points to the Arrhenius equation, the activation energy for oxygen self-diffusion is 6.33 eV. Another indirect method to calculate the activation energy is based on $0.5E_{f,\mathrm{O-FP}} + E^m_{\mathrm{V_O}}$, where $E_{f,\mathrm{O-FP}}$ is the formation energy of the oxygen Frenkel pair and $E^m_{\mathrm{V_O}}$ is the migration energy of oxygen vacancy \cite{murch1987oxygen,arima2010molecular}. In ThO$_2$ with the current potential, it was calculated that $E_{f,\mathrm{O-FP}}=5.89$ eV \cite{cooper2014many} and $E^m_{\mathrm{V_O}}=0.78$ eV (lowest barrier along [001] direction \cite{ghosh2016computational}), leading to 3.725 eV. This value is much lower than the direct method, suggesting that within this temperature range, oxygen diffusion may take high energy paths due to strong anharmonicity, e.g., the oxygen vacancy migration barrier along [110] direction is calculated to be 4.77 eV \cite{ghosh2016computational}. It should be noted that, the oxygen self-diffusion activation energy in ThO$_2$ deduced from experiments is around 2 eV (see Introduction part); this could be ascribed to the intrinsic defects that already exist in the samples for the measurements.  
\subsection{Dislocation}

With the setup of a dislocation pair, we calculate the oxygen diffusion coefficients as a function of the distance from the dislocation core ($r$). It is worth noting that dislocation migration is mostly avoided in our simulations, as the dislocation pair is at a relatively stable position from the equilibration stage (see Methods). To ensure that dislocation migration is avoided during the evaluation of MSD, dislocation behavior is monitored throughout the simulations, and only the non-moving dislocation instances are used for MSD statistics. The reason is that as the dislocation migrates, the oxygen displacements are found to populate across the crossed slide plane.

\begin{figure}[!ht]
	\centering
	\includegraphics[width=0.95\textwidth]{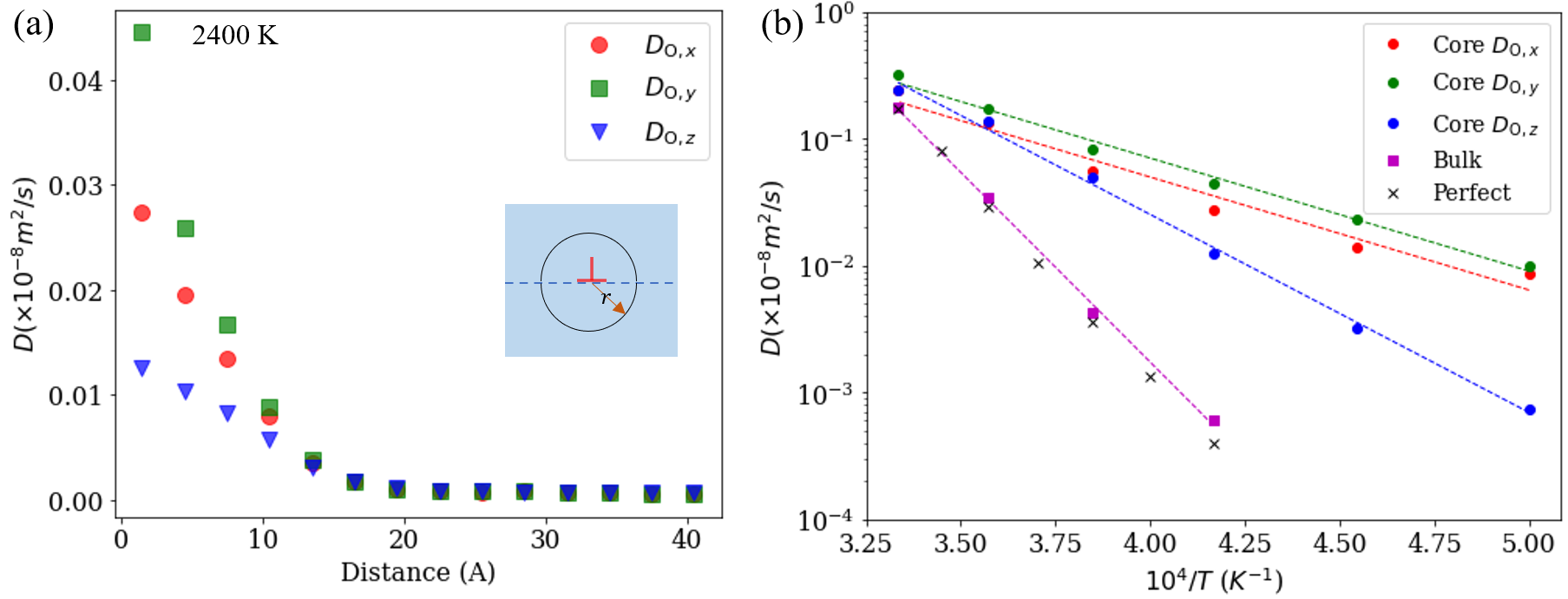}
	\caption{(a) Oxygen diffusion coefficient a function of distance from the dislocation core at 2400 K. (b) $D_\mathrm{O}$ versus the reciprocal of temperature for dislocation core and bulk oxygen atoms. Results from the perfect lattice calculation are provided as a reference.}
	\label{fig:dislocation_D}
\end{figure}
Figure \ref{fig:dislocation_D}a displays $D_{\mathrm{O},\alpha}$ as a function of $r$ at 2400 K (similar profiles are found for other temperatures). Oxygen diffusion is found significantly enhanced in the core region, especially along the dislocation line $\mathbf{\xi}$ ($D_\mathrm{O,y}$), which is attributed to pipe diffusion. Along $\mathbf{\xi \times b}$ on the dislocation half-plane, oxygen diffusivity ($D_\mathrm{O,z}$) is least enhanced. Figure \ref{fig:dislocation_D}b plots $D_\mathrm{O}$ versus temperature, comparing the values at the dislocation core and the bulk (regions $\sim4$ nm away from the dislocation core). $D_\mathrm{O}$ of oxygen atoms close to the dislocation core is significantly enhanced along all axes, compared to the bulk values, and it also shows higher values on the $x$-$y$ slide plane than in the perpendicular direction ($z$) at the low-temperature end. By fitting to the Arrhenius curve, the activation energies are obtained, 1.77 eV, 1.77 eV, and 3.09 eV along the $x$-, $y$-, and $z$-axis, respectively; it suggests that the higher $D_\mathrm{O,y}$ than $D_\mathrm{O,x}$ is due to the pre-factor instead of activation energy.       
\begin{figure}[!ht]
	\centering
	\includegraphics[width=0.95\textwidth]{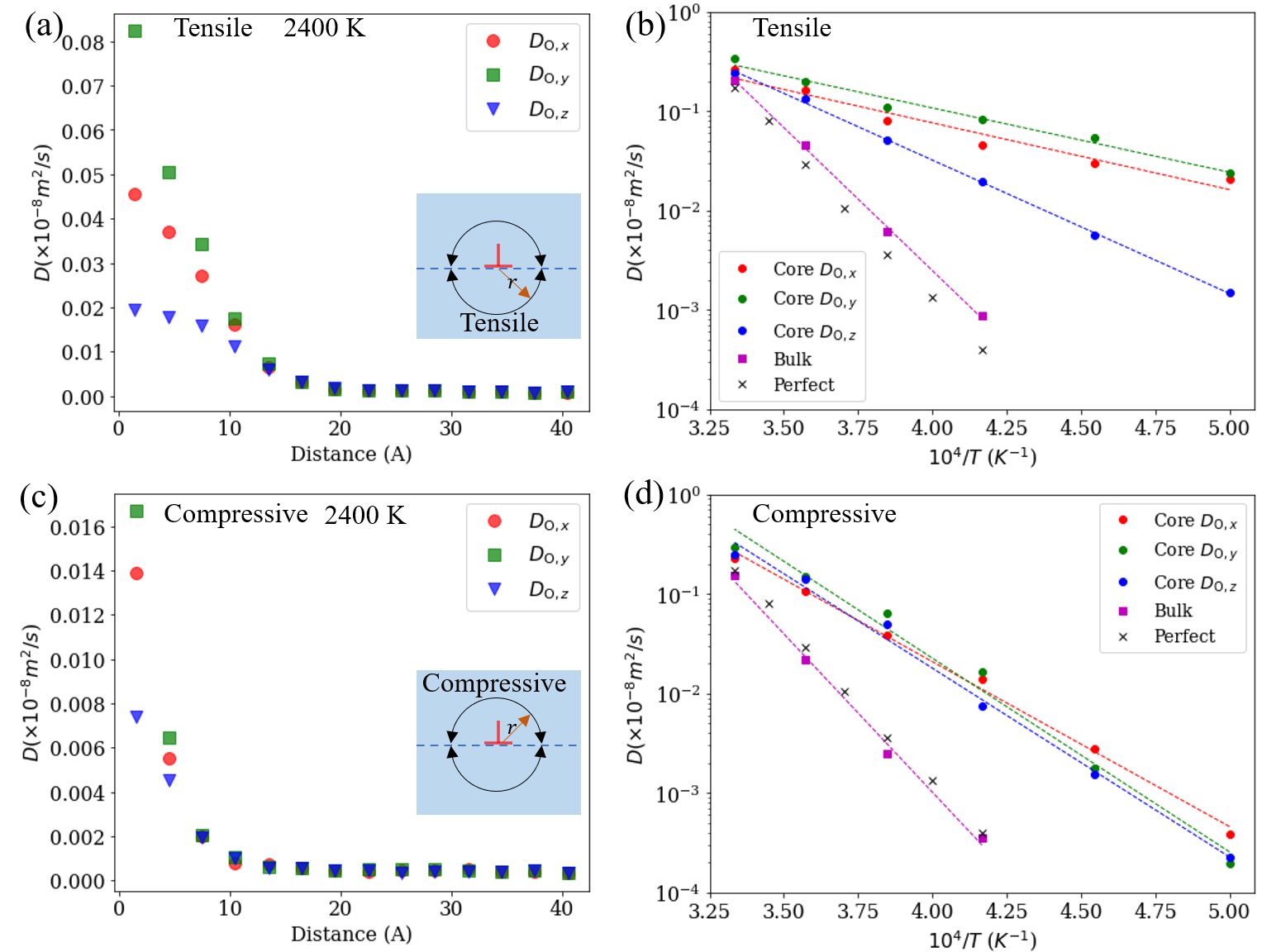}
	\caption{Oxygen diffusion coefficient a function of $r$ at 2400 K in the tensile field (a) and compressive field (c). $D_\mathrm{O}$ versus the reciprocal of temperature in the tensile field (b) and compressive field (d).  In (b) and (d), results from the perfect lattice calculation are provided as a reference.}
	\label{fig:dislocation_anisotropy}
\end{figure}

The local strain field is expected to cause anisotropy in oxygen diffusivity near the dislocation. In the study of UO$_2$ by Murphy et al. \cite{murphy2014pipe}, it was suggested that oxygen diffusion increases in the tensile region but is reduced in the compressive region, leading to a general increase in oxygen diffusivity. In another study of strained CeO$_2$, the enhancement of oxygen diffusivity in the bulk under tension surpasses the decrease in diffusivity when an equivalent amount of train is imposed \cite{rushton2013effect}. To reveal the effect in ThO$_2$, we divide the space into tensile and compressive regions, and separately calculate the oxygen diffusivity as a function of $r$ and $T$. FIGs. \ref{fig:dislocation_anisotropy}a and \ref{fig:dislocation_anisotropy}c show the spatially varying $D_\mathrm{O}$, indicating that $D_\mathrm{O}$ in the compressive region is much lower than in the tensile region. As the self-diffusion activation energy includes defect formation energy, the phenomenon could be attributed to a higher oxygen defect formation energy in the compressive region than that in the tensile region. Furthermore, the decay of $D_\mathrm{O}$ is faster in the compressive region, suggesting that the impact zone of the compressive elastic strain field is more localized than the tensile field. Furthermore, we compare the temperate dependent $D_\mathrm{O}$ at the dislocation core and bulk part for the tensile and compressive regions (FIGs. \ref{fig:dislocation_anisotropy}c and \ref{fig:dislocation_anisotropy}d). In the tensile field, the following sequence is found in a broad temperature range, $D_\mathrm{O,y} > D_\mathrm{O,x} \gg D_\mathrm{O,z}$, while in the compressive field, $D_\mathrm{O,x} \approx D_\mathrm{O,y} \approx D_\mathrm{O,z}$ with an average activation energy of 3.8 eV. Note that both the compressive and tensile regions close to the dislocation core have enhanced oxygen diffusion, which could be attributed to the strong interaction between oxygen point defects and the dislocation core. Finally, by comparing the bulk values (show convergence in $D_\mathrm{O}$, e.g., FIGs. \ref{fig:dislocation_anisotropy}a and \ref{fig:dislocation_anisotropy}c) with those based on the perfect lattice, there is a slight deviation due to the long-range elastic strain field: the compressive strain causes lower oxygen diffusivity, while the tensile strain causes higher oxygen diffusivity, consistent with the conclusion in CeO$_2$ \cite{rushton2013effect}. 

\begin{figure}[!ht]
	\centering
	\includegraphics[width=0.5\textwidth]{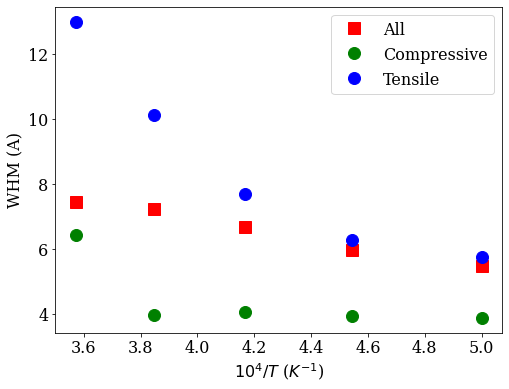}
	\caption{Dislocation impact zone $L_\mathrm{D}$ as a function of the reciprocal of temperature under three conditions. See text for definition.}
	\label{fig:dislocation_WHM}
\end{figure}

To quantify the impact zone of the dislocation for localized oxygen transport, we define a variable $L_\mathrm{D}$ as the zone radius (origin at dislocation core) at half maximum of $D_\mathrm{O}$ at the dislocation core. Figure \ref{fig:dislocation_WHM} shows three sets of analyses for the tensile, compressive, and combined regions as a function of the reciprocal of temperature. It can be seen that the impact zone noticeably increases with temperature in the tensile region, while it remains almost constant for the compressive region. Quantitatively, the dislocation-affected oxygen diffusion region is in the nanometer range from the core.

\subsection{GB}
\begin{figure}[!ht]
	\centering
	\includegraphics[width=0.95\textwidth]{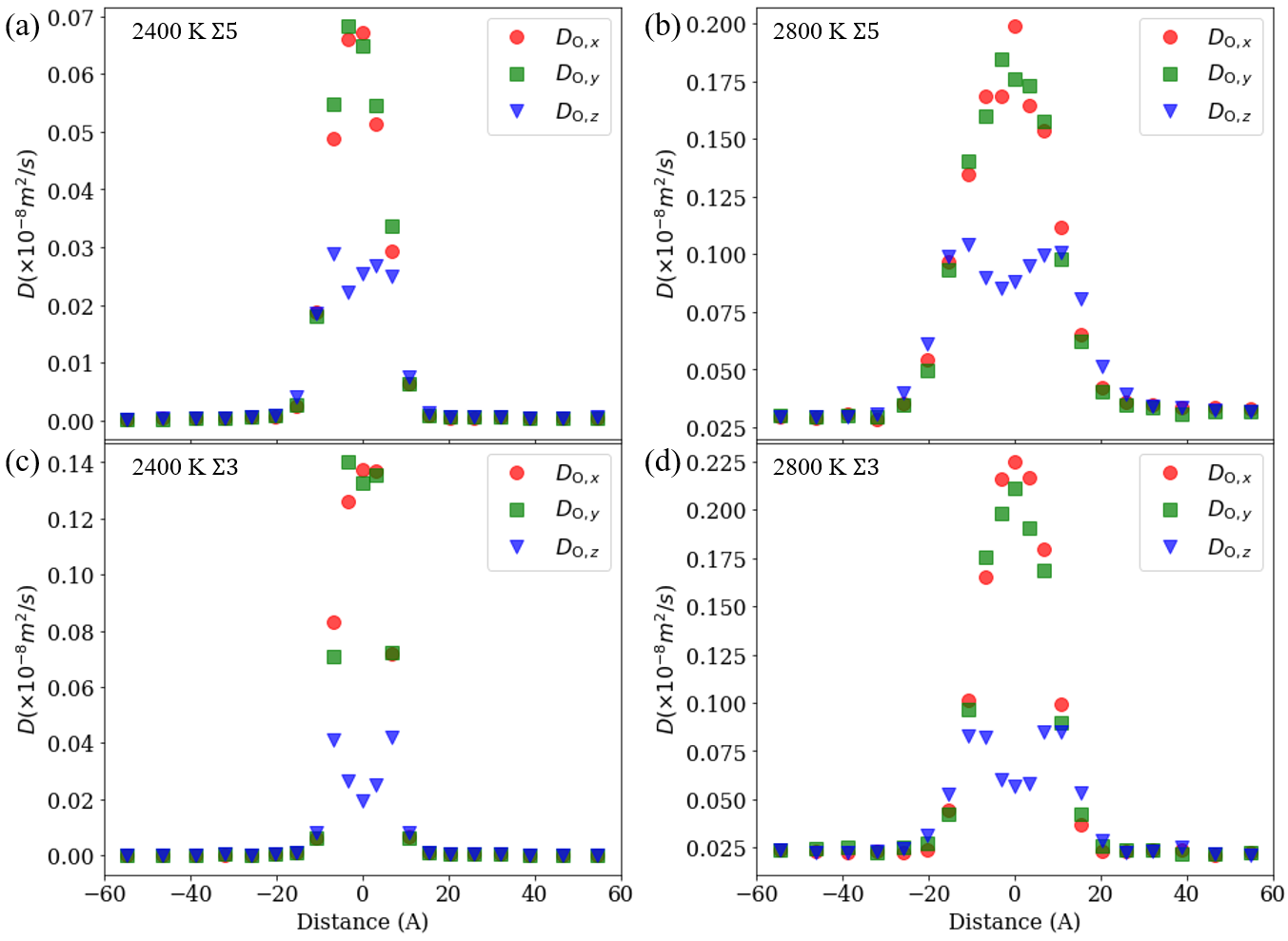}
	\caption{Oxygen diffusion coefficient a function of distance from GB at 2400 K and 2800 K for $\Sigma 3$ and $\Sigma 5$ GBs.}
	\label{fig:GB_D}
\end{figure}

As the considered GBs are symmetric tilt, the diffusion profile should be symmetric with respect to the GB plane, as evidenced in Figure \ref{fig:GB_D}, which displays the profiles of $D_\mathrm{O}$ across the GBs at 2400 K and 2800 K. For both GB types, oxygen diffusion is highest at the GB regions. By distinguishing the GB plane and the normal directions, it can be seen that diffusion is approximately isotropic on the GB plane ($D_\mathrm{O,x}$ and $D_\mathrm{O,y}$), but relatively low along the normal direction of the GB plane $D_\mathrm{O,z}$. It should be noted that in $D_\mathrm{O,z}$, there is a slight dip in the GB core region. This observation is due to the diffusion of nearby oxygen atoms to GB and subsequent confined oxygen migration within the GB plane; hence, MSD along $z$-direction is relatively lower for the oxygen atoms initially within the GB region than that in the neighboring regions.    

\begin{figure}[!ht]
	\centering
	\includegraphics[width=0.95\textwidth]{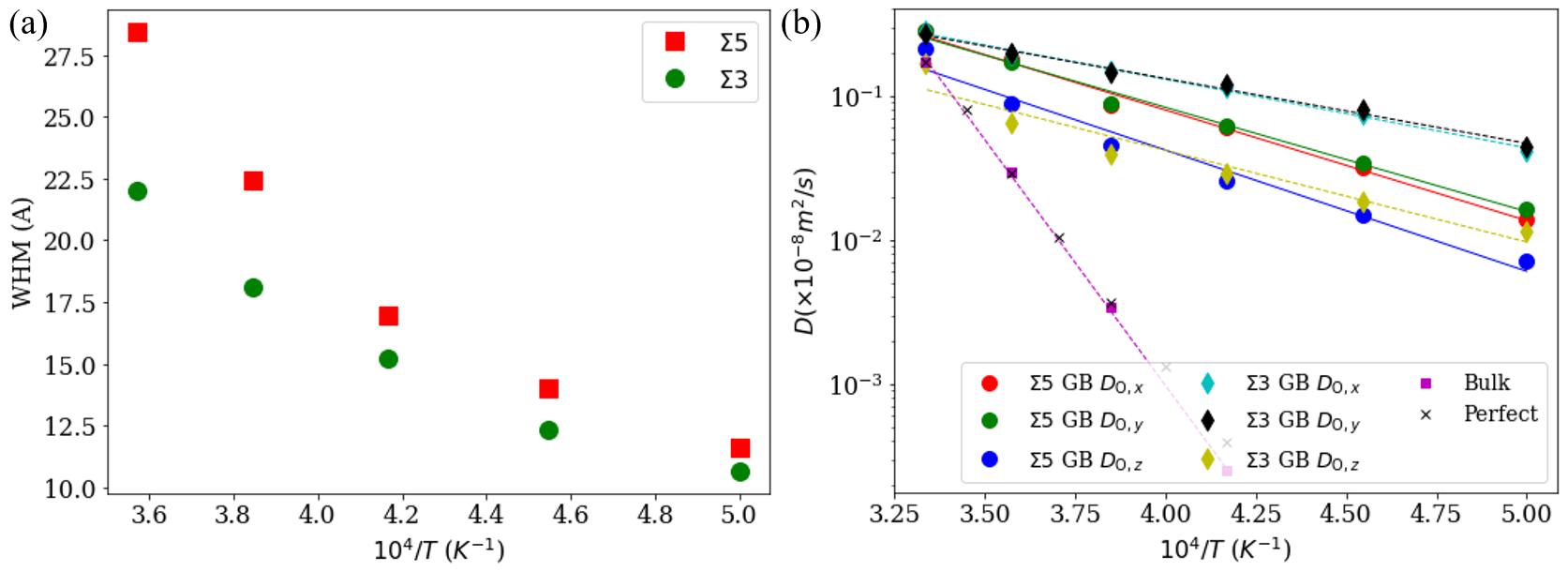}
	\caption{GB impact zone $L_\mathrm{GB}$ (a) and Oxygen diffusion coefficient at GB and bulk (b), as a function of the reciprocal of temperature. In (b), results from the perfect lattice calculation are provided as a reference.}
	\label{fig:GB_WHM}
\end{figure}

Figure \ref{fig:GB_D} also signifies the fact that the enhanced oxygen diffusivity is localized near the GB region, and the diffusivity quickly decays with distance. Here, the impact zone is quantified similarly by the $L_\mathrm{GB}$ defined by the zone width at half maximum of $D_\mathrm{O}$ evaluated at GBs. Figure \ref{fig:GB_WHM}a summarizes the results for the two GBs as a function of temperature. With increasing temperature, the impact zone increases for both GBs. The major impact range is around several nanometers, while $\Sigma 5$ has slightly higher widths than $\Sigma 3$ at all temperatures considered.   
 
With varying temperatures, we compute oxygen diffusivity at GB regions along each axis. Note that the axes correspond to different crystalline directions for both GBs and the perfect lattice, however, comparing the bulk values which evaluated 6 nm away from GB, with the perfect lattice values (Figure \ref{fig:GB_WHM}b), it can be concluded that oxygen diffusion is isotropic. Furthermore, it can be seen that diffusivity is consistently higher on GB plane than that along the plane normal. Furthermore, $D_\mathrm{O,x}$ and $D_\mathrm{O,y}$ nearly overlap for both GBs, yielding 1.48 eV for $\Sigma 5$ and 0.92 eV for $\Sigma 3$ for the activation energies based on Arrhenius fitting. Apparently, oxygen diffusivity at GB is dependent on the structure of GB, and $\Sigma 3$ GB plane possesses a high oxygen diffusivity than $\Sigma 5$. In $\Sigma 5(310)$, the open space in GB can equivalently be perceived as a high concentration of vacancies; if the vacancy-mediated diffusion mechanism is assumed dominant, the activation energy should be close to the oxygen migration energy. However, previous calculations of oxygen vacancy migration energy in the bulk give 0.78 eV \cite{colbourn1983calculated} and 0.64 eV \cite{he2022dislocation}, much lower than the fitted value 1.48 eV for $\Sigma 5(310)$ here. This discrepancy is presumably due to the contribution from the formation energy of the oxygen Frenkel pair in GB with weakened bonds at high temperatures. 

It appears counter-intuitive that $\Sigma 3$ GB exhibits a higher diffusivity than $\Sigma 5(310)$, as $\Sigma 5(310)$ has a large open space while $\Sigma 3(111)$ is compact (Figure \ref{fig:GB}). Note that, in UO$_2$, $\Sigma 3(111)$ GB was also found to exhibit strong oxygen diffusion \cite{williams2015atomistic}. For comparison of GB properties, we calculate the GB energy by ($E_{\mathrm{GB}}-E_0)/(2A)$, where $E_{\mathrm{GB}}$ is system potential energy with GB, $E_0$ is the perfect lattice potential energy with the same number of atoms, and $A$ is $x$-$y$ cross-section area, which leads to $E(\Sigma 5(310))=2.26~ \mathrm{J/m^2}$ and $E(\Sigma 3(111))=1.04~ \mathrm{J/m^2}$. It means that $\Sigma 3(111)$ GB is more energy favorable. In $\Sigma 3(111)$ GB, there are no inherent vacancies and interstitials in the ground state (Figure \ref{fig:GB}b). The high oxygen diffusivity and low activation energy suggest that both the oxygen Frenkel pair formation energy and the oxygen defect migration energy in $\Sigma 3(111)$ GB are low. Given that i) oxygen interstitials have a higher migration barrier than vacancies (0.78 eV \cite{colbourn1983calculated} and 0.64 eV \cite{he2022dislocation}) in ThO$_2$ \cite{he2022dislocation,colbourn1983calculated}, and ii) $\Sigma 3(111)$ GB exhibits 0.92 eV activation energy, lower than $\Sigma 5(310)$ with 1.48 eV, it is possible that other low-barrier oxygen diffusion mechanisms beyond the regular oxygen interstitial or vacancy mechanisms are activated within the $\Sigma 3(111)$ GB structure.

\subsection{Void}

It is known that the surface curvature can modify the diffusion-reaction energy barriers. For example, previous studies on the diffusion of adatom on the carbon nanotube surface indicate that positive curvature of the surface increases the diffusion barrier \cite{shu2001curvature}. 

\begin{figure}[!ht]
	\centering
	\includegraphics[width=0.95\textwidth]{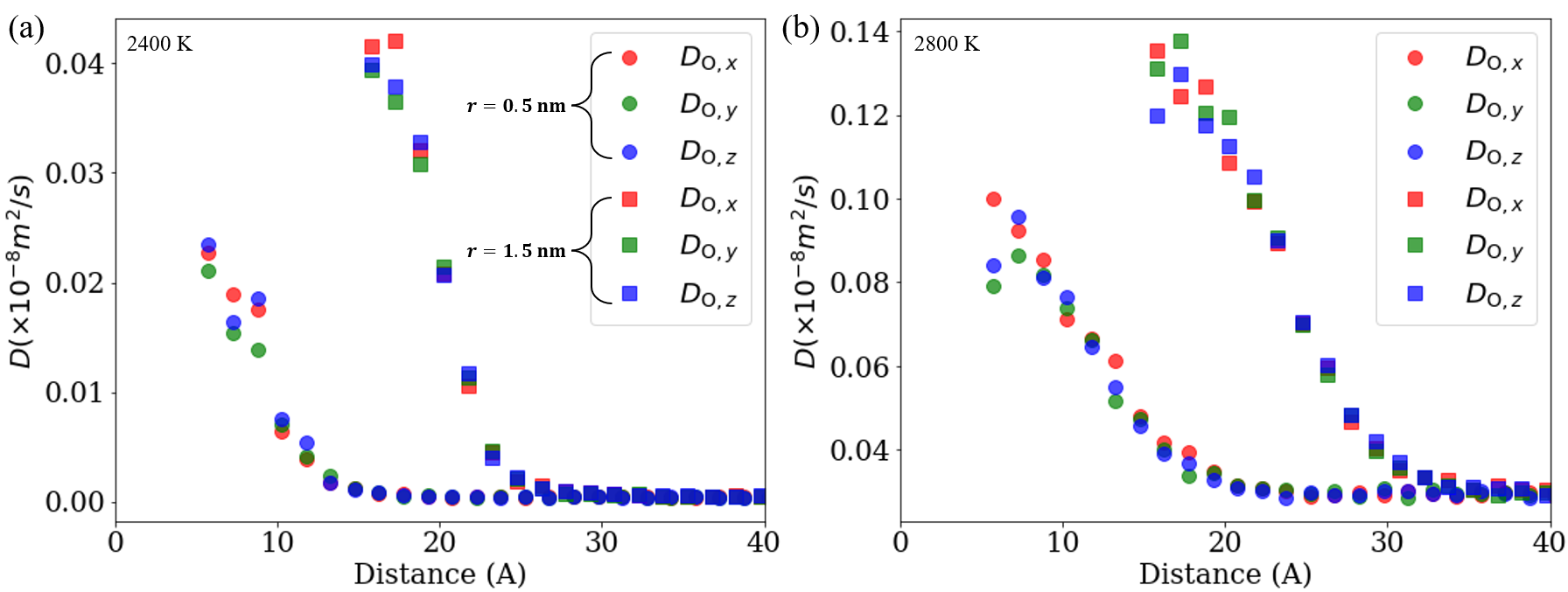}
	\caption{Oxygen diffusion coefficient a function of distance from the void center at 2400 K (a) and 2800 K (b).}
	\label{fig:void_comparison}
\end{figure}

Figure \ref{fig:void_comparison} compares $D_\mathrm{O}$ for two void sizes at 2400 K and 2800 K, where the distance at 0 \AA~denotes the center of the void. It can be seen that $D_\mathrm{O}$ is significantly enhanced at the void surface. Since the void is not faceted upon construction and remains spherical during the simulation process, there is no directional dependence in $D_\mathrm{O}$. Therefore, to extract the diffusion activation energy, the three components are averaged to increase the statistical quality of the fitting. Figure \ref{fig:void_comparison} plots ln$D_\mathrm{O}$ vs. $1/T$ for atoms at the interface, the values obtained in regions far from the void (i.e., bulk) are also provided for comparison, which reproduces the values based on the perfect lattice. Arrhenius fitting of the data points within 3 \AA~from the void surface yields the activation energies of oxygen diffusion at the void surface: $E_{a,\mathrm{vs}}(r=0.5\, \mathrm{nm}) = 1.84\, \mathrm{eV}$ and $E_{a,\mathrm{vs}}(r=1.5\, \mathrm{nm}) = 1.50\,\mathrm{eV}$.  

Further, small void sizes exhibit lower $D_\mathrm{O}$ on the void surface compared to large sizes. Theoretically, we can expect that the diffusion activation energy is modified by surface pressure $P=2\gamma /r$ resultant from the curvature, where $\gamma$ is the surface energy and $r$ is the void radius, i.e., $D\propto \mathrm{exp}[-(E_a+P\Delta V)/k_BT]$ \cite{butrymowicz1973diffusion}, where $\Delta V$ is the activation volume. In principle, there can be a nontrivial relationship between $\Delta V$ and temperature \cite{chroneos2015modeling,sarlis2016pressure}; for simplicity, it is approximated to be the oxygen atomic volume at 2000 K, i.e., $\Delta V\approx \Omega_\mathrm{O} = 15.4\, \mathrm{A}^3$ \cite{chroneos2015modeling}. By assuming constant surface energy, we can deduce the oxygen surface diffusion activation energy $E_{a,\mathrm{s}}$ and $\gamma$ based on two void sizes, i.e.,
\begin{align*} 
E_{a,\mathrm{s}} + \frac{2\gamma\Omega_\mathrm{O}}{0.5\, \mathrm{nm}} &=  E_{a,\mathrm{vs}}(r=0.5\, \mathrm{nm}) \\ 
E_{a,\mathrm{s}} + \frac{2\gamma\Omega_\mathrm{O}}{1.5\, \mathrm{nm}} &=  E_{a,\mathrm{vs}}(r=1.5\, \mathrm{nm})
\end{align*}
which yields $E_{a,\mathrm{s}}=1.33\,\mathrm{eV}$ and $\gamma = 1.31\,\mathrm{J/m^2}$. This surface energy should be perceived as an average quantity, as the void is spherical rather than faceted. In ThO$_2$, surface energies of low-index planes including (100), (110), and (111) planes have been quantified by atomistic methods  based on empirical potentials \cite{behera2012atomistic,benson1963calculation} and quantum mechanical treatment \cite{skomurski2008corrosion}, where the latter leads to $\gamma_{(111)}=0.72\,\mathrm{J/m^2}$, $\gamma_{(110)}=1.30\,\mathrm{J/m^2}$, and $\gamma_{(100)}=1.75\,\mathrm{J/m^2}$. The current fitted value is in the range of previous studies. Similarly, the surface diffusion activation energy $E_{a,\mathrm{s}}=1.33\,\mathrm{eV}$ is not limited to a specific plane, and should be treated as an average over general surfaces. 

\begin{figure}[!ht]
	\centering
	\includegraphics[width=0.95\textwidth]{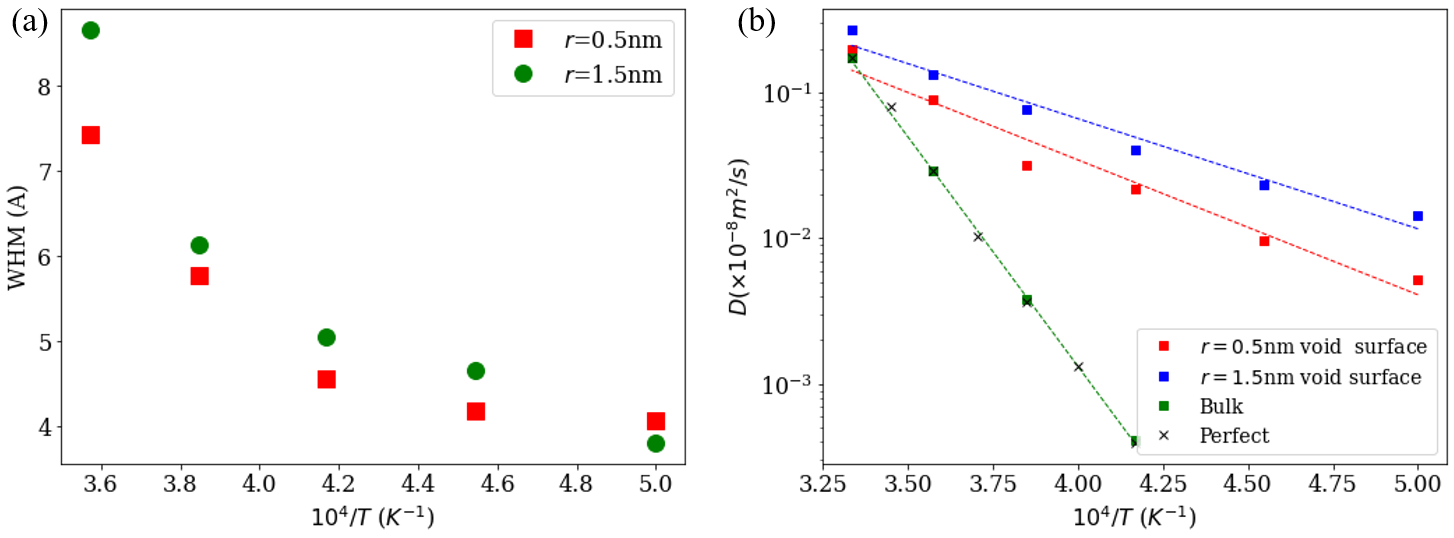}
	\caption{(a) Void impact zone $L_\mathrm{Void}$ for $r=0.5\, \mathrm{nm}$ and $r=1.5\, \mathrm{nm}$ voids. (b) Oxygen diffusion coefficient as a function of temperature at the void surface, bulk region for $r=0.5\, \mathrm{nm}$ and $r=1.5\, \mathrm{nm}$ systems. Results from the perfect lattice calculation are provided as a reference.}
	\label{fig:void_WHM}
\end{figure}

Finally, we compute the impact zone of a void on oxygen diffusion via a variable $L_\mathrm{Void}$ defined as the surface width at half the maximum of $D_\mathrm{O}$ evaluated at the void surface. Figure \ref{fig:void_WHM}a summarizes $L_\mathrm{Void}$ as a function of $1/T$, indicating that the impact zone strongly depends on the temperature but appears less sensitive to the void size. The entire width of the impact zone is around 1$\sim$2 nm, which is generally smaller than the case of dislocation and GBs.        

\section{Summary}
\begin{table}[h!]
\centering
\caption{Summary of activation energies (eV) for oxygen diffusion Values are obtained from most affected regions (i.e., dislocation core, GB plane, and void surface) of the extended defects. Refer to FIGs. \ref{fig:dislocation} and \ref{fig:GB} regarding the directions $x$, $y$ and $z$. }
\begin{tabular}{l|l|l|l|l|l|l}
\hline
 & \vtop{\hbox{\strut Edge}\hbox{\strut dislocation}} & \vtop{\hbox{\strut GB}\hbox{\strut $\Sigma 5(310)$}} & \vtop{\hbox{\strut GB}\hbox{\strut $\Sigma 3(111)$}} & \vtop{\hbox{\strut Void}\hbox{\strut $r=0.5$ nm}}          & \vtop{\hbox{\strut Void}\hbox{\strut $r=1.5$ nm}}           & \vtop{\hbox{\strut Perfect}\hbox{\strut lattice}}  \\ \hline
$x$         &     1.77             &  1.48   &  0.92   & \multirow{3}{*}{} & \multirow{3}{*}{} & \multirow{3}{*}{} \\ \cline{1-4}
$y$         &     1.77             &  1.48   &  0.92   &     1.84              &      1.50             &     6.33              \\ \cline{1-4}
$z$         &     3.08             &  N/A   &   N/A  &                   &                   &                   \\ \hline
\end{tabular}

\label{table:1}
\end{table}
 
By quantifying oxygen diffusion and the activation energy, it can be concluded that in pure ThO$_2$, various extended defects can enhance oxygen diffusion, although the boosting factors vary depending on the defect characteristics. TABLE \ref{table:1} summarizes the activation energy. For the 1D edge dislocation, pipe diffusion is obvious; the strain field, particularly near the core, significantly enhances oxygen diffusion, even in the compressive field, because of the strong interaction between oxygen defects with the dislocation core. Nevertheless, in the long-range elastic field, compressive strain leads to reduced oxygen diffusion than stress-free lattice, while tensile strain causes increased diffusion. For 2D defect GBs, the activation energy depends on the GB character, as also seen in UO$_2$ studies \cite{arima2010molecular,vincent2009self,williams2015atomistic}. However, it is not yet clear why $\Sigma 3(111)$ exhibits stronger oxygen diffusion than $\Sigma 5(310)$, where the latter has a large open space compared to $\Sigma 3(111)$. It is suspected that in  $\Sigma 3(111)$ GB, there are other oxygen diffusion mechanisms involved, which warrants further investigation. For 3D voids, oxygen diffusion is affected by local curvature/pressure, and smaller curvature leads to faster diffusion. It suggests that in ThO$_2$ at certain porosity (95\%), oxygen may diffuse significantly along the void surfaces. It remains to be confirmed whether this conclusion applies to pressurized bubbles in irradiated fuels.           

In the current treatment, we assign the atoms into slabs/shells at the initial time step. Close to the defect sinks, strong oxygen diffusion is identified, and an impact zone width of these defects is computed. Notably, this zone width may also be perceived as a metric for defect capture radius to quantify the defect sink strength for oxygen defects. Atomistically, there could be two bi-directional processes to explain the impact zone with enhanced oxygen transport: i) oxygen defects at the defect sink migrate toward the bulk, increasing local oxygen diffusivity, and ii) oxygen defects are generated near the defect sink and migrate toward the sink where oxygen diffusion is high. Therefore, such zone width should depend on temperature and defect characteristics, as confirmed by the results. 

These results indicate that extended defects-assisted diffusion can play a significant role in oxygen transport, especially at low temperatures. It should be noted that when comparing the data among those multidimensional defects, GB, particularly $\Sigma 3(111)$ shows the strongest impact to mediate the overall oxygen transport. Since $\Sigma 3(111)$ is also a low-energy GB type, its contribution to total oxygen diffusivity could be significant in polycrystalline ThO$_2$. 
 

 \bibliographystyle{elsarticle-num} 
 \bibliography{cas-refs}





\end{document}